\documentclass[12pt]{iopart}%
\usepackage{amsfonts}
\usepackage{amssymb}
\usepackage{graphicx}%

\begin{document}
\title[Space Time Foam]{Multigravity and Space Time Foam}
\author{Remo Garattini}

\begin{abstract}
We consider a multigravity approach to spacetime foam. As an application we
give indications on the computation of the cosmological constant, considered
as an eigenvalue of a Sturm-Liouville problem. A variational approach with
Gaussian trial wave functionals is used as a method to study such a problem.
We approximate the equation to one loop in a Schwarzschild background and a
zeta function regularization is involved to handle with divergences. The
regularization is closely related to the subtraction procedure appearing in
the computation of Casimir energy in a curved background. A renormalization
procedure is introduced to remove the infinities together with a
renormalization group equation.

\end{abstract}

\address{Universit\`{a} degli Studi di Bergamo, Facolt\`{a} di Ingegneria,\\ Viale
Marconi 5, 24044 Dalmine (Bergamo) ITALY.\\INFN - sezione di Milano, Via Celoria 16, Milan, Italy}
\ead{remo.garattini@unibg.it}\bigskip

Despite of its recent interest, the term \textquotedblleft%
\textit{Multigravity}\textquotedblright\ is not completely new. Indeed, in the
early seventies, some pioneering works appeared under the name
\textquotedblleft\textit{strong gravity}\textquotedblright\ or
\textquotedblleft\textit{f-g theory}\textquotedblright\cite{ISS} as a
tentative to describe a sector of hadronic physics where a massive spin-2
field (the \textit{f-meson} with Planck mass $M_{f}\sim1$ $GeV$) plays a
dominant role. Multigravity coincides with \textquotedblleft\textit{strong
gravity}\textquotedblright\ or \textquotedblleft\textit{f-g theory}%
\textquotedblright\ when the number of gravitational fields involved is
exactly $2$ (\textquotedblleft\textit{bigravity}\textquotedblright). In this
paper, we would like to use the Multigravity idea as a model of space-time
foam\cite{Wheeler} to compute the cosmological constant. Such a computation
has been done looking at the foam as a large $N$ composition of Schwarzschild
wormholes\cite{Remo}. Nevertheless, the Multigravity framework seems to be
more appropriate for such a computation. We recall that there exists a very
large discrepancy between the recent estimates on the cosmological constant,
which approximately are of the order of $10^{-47}GeV^{4}$, while a crude
estimate of the Zero Point Energy (ZPE) of some field of mass $m$ with a
cutoff at the Planck scale gives $E_{ZPE}\approx10^{71}GeV^{4}$ with a
difference of about 118 orders\cite{Lambda}. Let us see how to use
multigravity, to approach such a problem. To this purpose we begin with the
following action involving $N$ massless gravitons without matter
fields\cite{DamourKogan}%
\begin{equation}
S_{0}=\sum_{i=1}^{N}S\left[  g_{i}\right]  =\sum_{i=1}^{N}\frac{1}{16\pi
G_{i}}\int d^{4}x\sqrt{-g_{i}}\left[  R\left(  g_{i}\right)  -\Lambda
_{i}\right]  ,
\end{equation}
where $\Lambda_{i}$ and $G_{i}$ are the cosmological constant and the related
Newton constant corresponding $i^{th}$ universe, respectively. Generally
speaking, the total action should be of the form%
\begin{equation}
S_{tot}=\sum_{i=1}^{N}S\left[  g_{i}\right]  +\lambda S_{int}\left(
g_{1},g_{2},\ldots,g_{N}\right)  .
\end{equation}
When $\lambda\rightarrow0,$the $N$ world are non-interacting. This will be the
context we are going to examine. In this way, the action $S_{0}$ describes a
gas of gravitons. Consider for simplicity the case $N=1$ and the related
Einstein field equations%
\begin{equation}
R_{\mu\nu}-\frac{1}{2}g_{\mu\nu}R^{\left(  4\right)  }+\Lambda_{c}g_{\mu\nu
}=G_{\mu\nu}+\Lambda_{c}g_{\mu\nu}=0,
\end{equation}
where $G_{\mu\nu}$ is the Einstein tensor. If we introduce a time-like unit
vector $u^{\mu}$ such that $u\cdot u=-1$, then%
\begin{equation}
G_{\mu\nu}u^{\mu}u^{\mu}=\Lambda_{c}.
\end{equation}
This is simply the Hamiltonian constraint written in terms of equation of
motion, which is classical. However, the discrepancy between the observed
cosmological constant and the computed one is in its quantum version, that
could be estimate by the expectation value $\left\langle \Lambda
_{c}\right\rangle $. Since%
\begin{equation}
\frac{\sqrt{g}}{2\kappa}G_{\mu\nu}u^{\mu}u^{\mu}=\frac{\sqrt{g}}{2\kappa
}R+\frac{2\kappa}{\sqrt{g}}\left(  \frac{\pi^{2}}{2}-\pi^{\mu\nu}\pi_{\mu\nu
}\right)  =-\mathcal{H},
\end{equation}
where $R$ is the scalar curvature in three dimensions, we can write%
\begin{equation}
\frac{\left\langle \Lambda_{c}\right\rangle }{\kappa}=-\frac{1}{V}\left\langle
\int_{\Sigma}d^{3}x\mathcal{H}\right\rangle =-\frac{1}{V}\left\langle
\int_{\Sigma}d^{3}x\hat{\Lambda}_{\Sigma}\right\rangle ,
\end{equation}
where the last expression stands for%
\begin{equation}
\frac{1}{V}\frac{\int\mathcal{D}\left[  g_{ij}\right]  \Psi^{\ast}\left[
g_{ij}\right]  \int_{\Sigma}d^{3}x\mathcal{H}\Psi\left[  g_{ij}\right]  }%
{\int\mathcal{D}\left[  g_{ij}\right]  \Psi^{\ast}\left[  g_{ij}\right]
\Psi\left[  g_{ij}\right]  }=\frac{1}{V}\frac{\left\langle \Psi\left\vert
\int_{\Sigma}d^{3}x\hat{\Lambda}_{\Sigma}\right\vert \Psi\right\rangle
}{\left\langle \Psi|\Psi\right\rangle }=-\frac{\Lambda}{\kappa},\label{expect}%
\end{equation}
and where we have integrated over the hypersurface $\Sigma$, divided by its
volume and functionally integrated over quantum fluctuation with the help of
some trial wave functionals. Note that Eq.$\left(  \ref{expect}\right)  $ can
be derived starting with the Wheeler-De Witt equation (WDW) \cite{De Witt}
which represents invariance under \textit{time} reparametrization. Eq.$\left(
\ref{expect}\right)  $ represents the Sturm-Liouville problem associated with
the cosmological constant. The related boundary conditions are dictated by the
choice of the trial wavefunctionals which, in our case are of the Gaussian
type. Different types of wavefunctionals correspond to different boundary
conditions. Extracting the TT tensor contribution from Eq.$\left(
\ref{expect}\right)  $ approximated to second order in perturbation of the
spatial part of the metric into a background term, $\bar{g}_{ij}$, and a
perturbation, $h_{ij}$, we get
\begin{equation}
\hat{\Lambda}_{\Sigma}^{\bot}=\frac{1}{4V}\int_{\Sigma}d^{3}x\sqrt{\bar{g}%
}G^{ijkl}\left[  \left(  2\kappa\right)  K^{-1\bot}\left(  x,x\right)
_{ijkl}+\frac{1}{\left(  2\kappa\right)  }\left(  \triangle_{2}\right)
_{j}^{a}K^{\bot}\left(  x,x\right)  _{iakl}\right]  .\label{p22}%
\end{equation}
Here $G^{ijkl}$ represents the inverse DeWitt metric and all indices run from
one to three. The propagator $K^{\bot}\left(  x,x\right)  _{iakl}$ can be
represented as
\begin{equation}
K^{\bot}\left(  \overrightarrow{x},\overrightarrow{y}\right)  _{iakl}:=%
{\displaystyle\sum_{\tau}}
\frac{h_{ia}^{\left(  \tau\right)  \bot}\left(  \overrightarrow{x}\right)
h_{kl}^{\left(  \tau\right)  \bot}\left(  \overrightarrow{y}\right)
}{2\lambda\left(  \tau\right)  },\label{proptt}%
\end{equation}
where $h_{ia}^{\left(  \tau\right)  \bot}\left(  \overrightarrow{x}\right)  $
are the eigenfunctions of $\triangle_{2}$, whose explicit expression for the
massive case will be shown in the next section. $\tau$ denotes a complete set
of indices and $\lambda\left(  \tau\right)  $ are a set of variational
parameters to be determined by the minimization of Eq.$\left(  \ref{p22}%
\right)  $. The expectation value of $\hat{\Lambda}_{\Sigma}^{\bot}$ is easily
obtained by inserting the form of the propagator into Eq.$\left(
\ref{p22}\right)  $ and minimizing with respect to the variational function
$\lambda\left(  \tau\right)  $. Thus the total one loop energy density for TT
tensors becomes%
\begin{equation}
\frac{\Lambda}{8\pi G}=-\frac{1}{4V}%
{\displaystyle\sum_{\tau}}
\left[  \sqrt{\omega_{1}^{2}\left(  \tau\right)  }+\sqrt{\omega_{2}^{2}\left(
\tau\right)  }\right]  .\label{1loop}%
\end{equation}
The above expression makes sense only for $\omega_{i}^{2}\left(  \tau\right)
>0$, where $\omega_{i}$ are the eigenvalues of $\triangle_{2}$. If we fix our
attention on some particular background, for example the Schwarzschild
background, the spin 2 operator $\triangle_{2}$, simply becomes

The further step is the evaluation of Eq.$\left(  \ref{1loop}\right)  $. Its
contribution to the Spin-two operator for the Schwarzschild metric will be%
\begin{equation}
\left(  \triangle_{2}h^{TT}\right)  _{i}^{j}:=-\triangle_{S}\left(
h^{TT}\right)  _{i}^{j}+\frac{6}{r^{2}}\left(  1-\frac{2MG}{r}\right)  \left(
h^{TT}\right)  _{i}^{j}+2\left(  Rh^{TT}\right)  _{i}^{j}. \label{spin2}%
\end{equation}
$\triangle_{S}$ is the scalar curved Laplacian, whose form is%
\begin{equation}
\triangle_{S}=\left(  1-\frac{2MG}{r}\right)  \frac{d^{2}}{dr^{2}}+\left(
\frac{2r-3MG}{r^{2}}\right)  \frac{d}{dr}-\frac{L^{2}}{r^{2}} \label{slap}%
\end{equation}
and $R_{j}^{a}$ is the mixed Ricci tensor whose components are:
\begin{equation}
R_{i}^{a}=\left\{  -\frac{2MG}{r^{3}},\frac{MG}{r^{3}},\frac{MG}{r^{3}%
}\right\}  .
\end{equation}
This implies that the scalar curvature is traceless. We are therefore led to
study the following eigenvalue equation
\begin{equation}
\left(  \triangle_{2}h^{TT}\right)  _{i}^{j}=\omega^{2}h_{j}^{i} \label{p31}%
\end{equation}
where $\omega^{2}$ is the eigenvalue of the corresponding equation. In doing
so, we follow Regge and Wheeler in analyzing the equation as modes of definite
frequency, angular momentum and parity\cite{Regge Wheeler}. In particular, our
choice for the three-dimensional gravitational perturbation is represented by
its even-parity form%
\begin{equation}
\left(  h^{even}\right)  _{j}^{i}\left(  r,\vartheta,\phi\right)  =diag\left[
H\left(  r\right)  ,K\left(  r\right)  ,L\left(  r\right)  \right]
Y_{lm}\left(  \vartheta,\phi\right)  . \label{pert}%
\end{equation}
Defining reduced fields and passing to the proper geodesic distance from the
\textit{throat} of the bridge, the system $\left(  \ref{p31}\right)  $ becomes%

\begin{equation}
\left\{
\begin{array}
[c]{c}%
\left[  -\frac{d^{2}}{dx^{2}}+\frac{l\left(  l+1\right)  }{r^{2}}+m_{1}%
^{2}\left(  r\right)  \right]  f_{1}\left(  x\right)  =\omega_{1,l}^{2}%
f_{1}\left(  x\right) \\
\\
\left[  -\frac{d^{2}}{dx^{2}}+\frac{l\left(  l+1\right)  }{r^{2}}+m_{2}%
^{2}\left(  r\right)  \right]  f_{2}\left(  x\right)  =\omega_{2,l}^{2}%
f_{2}\left(  x\right)
\end{array}
\right.  \label{p34}%
\end{equation}
where we have defined $r\equiv r\left(  x\right)  $ and%
\begin{equation}
\left\{
\begin{array}
[c]{c}%
m_{1}^{2}\left(  r\right)  =U_{1}\left(  r\right)  =m_{1}^{2}\left(
r,M\right)  -m_{2}^{2}\left(  r,M\right) \\
\\
m_{2}^{2}\left(  r\right)  =U_{2}\left(  r\right)  =m_{1}^{2}\left(
r,M\right)  +m_{2}^{2}\left(  r,M\right)
\end{array}
\right.  .
\end{equation}
$m_{1}^{2}\left(  r,M\right)  \rightarrow0$ when $r\rightarrow\infty$ or
$r\rightarrow2MG$ and $m_{2}^{2}\left(  r,M\right)  =3MG/r^{3}$. Note that,
while $m_{2}^{2}\left(  r\right)  $ is constant in sign, $m_{1}^{2}\left(
r\right)  $ is not. Indeed, for the critical value $\bar{r}=5MG/2$, $m_{1}%
^{2}\left(  \bar{r}\right)  =m_{g}^{2}$ and in the range $\left(
2MG,5MG/2\right)  $ for some values of $m_{g}^{2}$, $m_{1}^{2}\left(  \bar
{r}\right)  $ can be negative. It is interesting therefore concentrate in this
range, where $m_{1}^{2}\left(  r,M\right)  $ vanishes when compared with
$m_{2}^{2}\left(  r,M\right)  $. So, in a first approximation we can write%
\begin{equation}
\left\{
\begin{array}
[c]{c}%
m_{1}^{2}\left(  r\right)  \simeq-m_{2}^{2}\left(  r_{0},M\right) \\
\\
m_{2}^{2}\left(  r\right)  \simeq+m_{2}^{2}\left(  r_{0},M\right)
\end{array}
\right.  ,
\end{equation}
where we have defined a parameter $r_{0}>2MG$ and $m_{0}^{2}\left(
r_{0},M\right)  =3MG/r_{0}^{3}$. The main reason for introducing a new
parameter resides in the fluctuation of the horizon that forbids any kind of
approach. It is now possible to explicitly evaluate Eq.$\left(  \ref{1loop}%
\right)  $ in terms of the effective mass. By adopting the W.K.B. method used
by `t Hooft in the brick wall problem\cite{tHooft}, we arrive at the following
relevant expression%
\begin{equation}
\rho_{i}\left(  \varepsilon\right)  =\frac{m_{i}^{4}\left(  r\right)  }%
{256\pi^{2}}\left[  \frac{1}{\varepsilon}+\ln\left(  \frac{\mu^{2}}{m_{i}%
^{2}\left(  r\right)  }\right)  +2\ln2-\frac{1}{2}\right]  , \label{zeta1}%
\end{equation}
$i=1,2$, where we have used the zeta function regularization method to compute
the energy densities $\rho_{i}$ and where we have introduced the additional
mass parameter $\mu$ in order to restore the correct dimension for the
regularized quantities. Such an arbitrary mass scale emerges unavoidably in
any regularization scheme. The energy density is renormalized via the
absorption of the divergent part (in the limit $\varepsilon\rightarrow0$) into
the re-definition of the bare classical constant $\Lambda$
\begin{equation}
\Lambda\rightarrow\Lambda_{0}+\Lambda^{div}=\Lambda_{0}+\frac{G}%
{32\pi\varepsilon}\left(  m_{1}^{4}\left(  r\right)  +m_{2}^{4}\left(
r\right)  \right)  .
\end{equation}
To remove the dependence on the arbitrary mass scale $\mu$, it is appropriate
to use the renormalization group equation. Therefore we impose that\cite{RGeq}%

\begin{equation}
\frac{1}{8\pi G}\mu\frac{\partial\Lambda_{0}^{TT}\left(  \mu\right)
}{\partial\mu}=\mu\frac{d}{d\mu}\rho_{eff}^{TT}\left(  \mu,r\right)  ,
\label{rg}%
\end{equation}
where $\rho_{eff}^{TT}\left(  \mu,r\right)  $ is the renormalized energy
density. Solving it we find that the renormalized constant $\Lambda_{0}$
should be treated as a running one in the sense that it varies provided that
the scale $\mu$ is changing%

\begin{equation}
\Lambda_{0}\left(  \mu,r\right)  =\Lambda_{0}\left(  \mu_{0},r\right)
+\frac{G}{16\pi}\left(  m_{1}^{4}\left(  r\right)  +m_{2}^{4}\left(  r\right)
\right)  \ln\frac{\mu}{\mu_{0}}.\label{lambdamu}%
\end{equation}
The final form for the cosmological constant is\cite{Remo1}%
\begin{equation}
\frac{\Lambda_{0}\left(  \mu_{0},r_{0}\right)  }{8\pi G}=-\frac{m_{0}%
^{4}\left(  r_{0},M\right)  }{128\pi^{2}}\ln\left(  \frac{m_{0}^{2}\left(
r_{0},M\right)  \sqrt{e}}{4}\right)  \label{lambdamu0a}%
\end{equation}
which has a minimum for%
\begin{equation}
\frac{m_{0}^{2}\left(  r_{0},M\right)  \sqrt{e}}{4\mu_{0}^{2}}=\frac{1}%
{\sqrt{e}}%
\end{equation}
with%
\begin{equation}
\frac{\Lambda_{0}\left(  \mu_{0},r\right)  }{8\pi G}=-\frac{\mu_{0}^{4}%
}{16e^{2}\pi^{2}}%
\end{equation}
We can now discuss the multigravity gas. For each gravitational field
introduce the following variables $\left(  N,N_{i}\right)  ^{\left(  k\right)
}$ and choose the gauge $N_{i}^{\left(  k\right)  }=0$, $\left(  k=1\ldots
N_{w}\right)  $. Define the following domain $D_{\Lambda}^{\left(  k\right)
}$%
\begin{equation}
\left\{  \Psi|\left[  \left[  \left(  2\kappa\right)  G_{ijkl}\pi^{ij}\pi
^{kl}-\frac{\sqrt{g}}{2\kappa}R\right]  ^{\left(  k\right)  }\Psi^{\left(
k\right)  }\left[  g_{ij}^{\left(  k\right)  }\right]  =-\frac{\sqrt
{g^{\left(  k\right)  }}}{\kappa^{\left(  k\right)  }}\Lambda_{c}^{\left(
k\right)  }\Psi^{\left(  k\right)  }\left[  g_{ij}^{\left(  k\right)
}\right]  \right]  \right\}  ,
\end{equation}
and assume the following assumption: $\exists\,$a covering of $\Sigma$ s.t.%
\begin{equation}
\Sigma=\bigcup\limits_{k=1}^{N_{w}}\Sigma_{k}\qquad\Sigma_{k}\cap\Sigma
_{j}=\emptyset\label{covering}%
\end{equation}
for $k\neq j$. Then the Eq.$\left(  \ref{expect}\right)  $ turns into%
\begin{equation}
\frac{1}{V_{\left(  k\right)  }}\frac{\int\mathcal{D}\left[  g_{ij}^{\left(
k\right)  }\right]  \Psi_{\left(  k\right)  }^{\ast}\left[  g_{ij}^{\left(
k\right)  }\right]  \int_{\Sigma_{k}}d^{3}x\hat{\Lambda}_{\Sigma_{k}}^{\left(
k\right)  }\Psi_{\left(  k\right)  }\left[  g_{ij}^{\left(  k\right)
}\right]  }{\int\mathcal{D}\left[  g_{ij}^{\left(  k\right)  }\right]
\Psi_{\left(  k\right)  }^{\ast}\left[  g_{ij}^{\left(  k\right)  }\right]
\Psi_{\left(  k\right)  }\left[  g_{ij}^{\left(  k\right)  }\right]  }%
=-\frac{\Lambda_{\left(  k\right)  }}{8\pi G_{\left(  k\right)  }%
},\label{expectk}%
\end{equation}
Each $\Sigma_{k}$ has topology $S^{2}\times R^{1}$. Therefore, the whole
physical space $\Sigma$ containing the energy density appears depicted as in
the following picture%
\begin{figure}
[pbh]
\begin{center}
\includegraphics[
height=2.5797in,
width=2.5797in
]%
{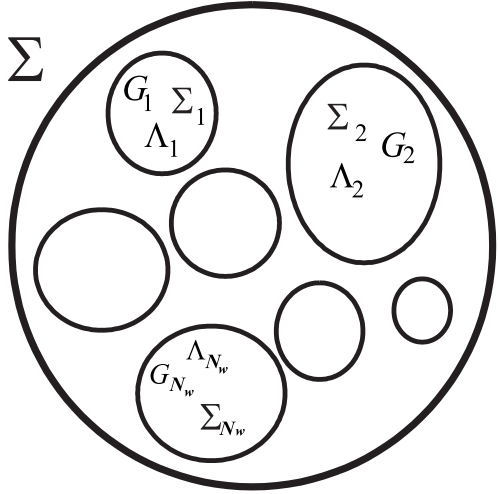}%
\caption{The space $\Sigma$ composed by the non overlapping spaces $\Sigma
_{k}$}%
\label{picture1}%
\end{center}
\end{figure}
A further simplification comes from the assumption that the different Newton's
constants are all equal. This leads to a model which is composed by $N_{w}$
copies of the same world\cite{Remo1} and on each copy the procedure contained
between Eq.$\left(  \ref{1loop}\right)  $ and Eq.$\left(  \ref{lambdamu0a}%
\right)  $ has to be repeated. Thus, the final evaluation of the
\textquotedblleft\textit{global}\textquotedblright\ cosmological constant can
be written%
\begin{equation}
\max\left\{  \frac{1}{V_{1}}\int_{\Sigma_{1}}d^{3}x\Lambda_{\Sigma_{1}%
}^{\left(  1\right)  }\right\}  +\ldots+\max\left\{  \frac{1}{V_{N_{w}}}%
\int_{\Sigma_{N_{w}}}d^{3}x\Lambda_{\Sigma_{N_{w}}}^{\left(  N_{w}\right)
}\right\}  =-\frac{\Lambda}{8\pi G},
\end{equation}
where $\Lambda_{\Sigma_{k}}^{\left(  k\right)  }$ is the eigenvalue obtained
evaluating Eq.$\left(  \ref{expectk}\right)  $ on each $\Sigma_{k}$. The
computation of the $\max$ is taken on each disjoint $\Sigma_{k}.$ Note that in
any case, the maximum of $\Lambda_{\Sigma_{k}}^{\left(  k\right)  }$
corresponds to the minimum of the energy density computed on the related
hypersurface\textbf{. }It is interesting also to note that the whole procedure
can be applied even in case of a massive graviton\cite{massive} with a term of
the form\cite{Remo2, Rubakov}%
\begin{equation}
S_{m}=\frac{m_{g}^{2}}{8\kappa}\int d^{4}x\sqrt{-\hat{g}}\left[  h^{ij}%
h_{ij}\right]  ,
\end{equation}
which is a particular sub-case of the Pauli-Fierz term\cite{PauliFierz}.

\end{document}